# Using Deep Learning to Examine the Association between the Built Environment and Neighborhood Adult Obesity Prevalence

Adyasha Maharana[1] and Elaine O. Nsoesie[2*]

1. Department of Biomedical Informatics and Medical Education, University of Washington, Seattle, WA 98195, United States
2. Institute for Health Metrics and Evaluation, University of Washington, Seattle, WA 98121, United States

*Corresponding author: en22@uw.edu

**ABSTRACT**

*More than one-third of the adult population in the United States is obese. Obesity has been linked to factors such as, genetics, diet, physical activity and the environment. However, evidence indicating associations between the built environment and obesity has varied across studies and geographical contexts. Here, we used deep learning and approximately 150,000 high resolution satellite images to extract features of the built environment. We then developed linear regression models to consistently quantify the association between the extracted features and obesity prevalence at the census tract level for six cities in the United States. The extracted features of the built environment explained 72% to 90% of the variation in obesity prevalence across cities. Out-of-sample predictions were considerably high with correlations greater than 80% between predicted and true obesity prevalence across all census tracts. This study supports a strong association between the built environment and obesity prevalence. Additionally, it also illustrates that features of the built environment extracted from satellite images can be useful for studying health indicators, such as obesity. Understanding the association between specific features of the built environment and obesity prevalence can lead to structural changes that could encourage physical activity and decreases in obesity prevalence.*

**INTRODUCTION**

The Global Burden of Disease study estimates that more than 603 million adults worldwide were obese in 2015 (1). In the United States, greater than one-third of the adult population is obese (2–4) and forty-six states have an estimated adult obesity rate of twenty-five percent or more (5). Obesity is a complex health issue that has been linked to a myriad of factors including, genetics, demographics, and behavior (6). Behavioral traits that encourage unhealthy food choices and a sedentary lifestyle have been associated with features in the social and built environment, which comprises both natural and modified elements of the physical environment. The built environment can influence health through exposure to chemical, biological and physical agents, and also through the availability of resources, such as, housing, activity and recreational spaces, and measures of community design (7).

Many studies have shown that certain features of the built environment can be associated with obesity and physical activity across different life stages (i.e., adolescents, adult, older adults) (8–26). For example, physical activity has been associated with housing density, mixed land-use, varied housing types, access to natural spaces and the availability of footpaths, and health clubs (27, 28). Research also supports the association between obesity and similar factors, including, walkability, land-use, sprawl, area of residence and access to resources (such as, recreational facilities and food outlets), level of deprivation and perceived safety (29–32). Proximity and access to natural spaces and sidewalks can lead to increased and regular physical activity, especially in urban areas (33–37). This association between walkable environments and physical activity has been noted across different geographical scales (9, 38). For example, a recent study using data from smartphones to assess physical activity globally suggested that inequalities in physical activity were predictive of obesity prevalence and features of the physical environment such as, walkability, lowered the activity inequality (38).

Despite these associations, there are noted inconsistencies across studies and geographical contexts on the association between specific features of the built environment and obesity prevalence (30, 39–41). These inconsistencies could be due to variations in measures and measurement tools across studies, making it difficult to assess and compare findings (31). Furthermore, the process of measuring these features can be costly, time consuming and subjective to human judgement and bias. Approaches that enable consistent measurement and allow for comparison across studies are needed. Furthermore, assessing and quantifying the impact of the built environment on obesity would be useful for selecting and implementing appropriate community-based interventions and prevention efforts (4, 31).

Here, we propose a method for comprehensively assessing the association between adult obesity prevalence and the built environment that involves extracting neighborhood physical features from high resolution satellite imagery using convolutional neural networks. We demonstrate our approach by predicting obesity prevalence at the census tract level for six cities – Los Angeles, California; Memphis, Tennessee; San Antonio, Texas; and Seattle, Tacoma and Bellevue, Washington – with varying obesity prevalence. We combined the data for Seattle, Tacoma and Bellevue into a single dataset (hereafter referred to as, Seattle) because each has a small number of census tracts. Our approach provides fine-grained associations between obesity prevalence and the built environment, and can also enable comparability across studies. Furthermore, our approach is scalable and relies on openly available data and computational tools.

**RESULTS**

We used a pre-trained convolutional neural network (CNN) to extract data (henceforth referred to as, features) that encode structures of the built environment present in satellite images (*Materials and Methods*). These features include, buildings, greenery, roads, water, land etc. (see Fig. 1). The extracted features were then used as predictors in a regression model, where the response variable was the obesity prevalence for each census tract. The features of the built environment explained 75% of the variation in obesity prevalence across all census tracts (n=1,695). Fig. 2 (a) displays the strong linear correlation between the actual obesity prevalence and the cross-validated model estimated prevalence.

To investigate whether this trend was consistent, we modeled the data for each region separately. Individually, our models explained 72%, 82%, 85% and 91% of the variation in neighborhood obesity prevalence for Los Angeles, San Antonio, Seattle and Memphis, respectively (Fig. 3 and 4). Los Angeles (Fig. 4(c)) has the weakest association because our approach underestimates the highest obesity prevalence, which is mostly clustered in the southern part of the city. However, our approach consistently presents a strong association between obesity prevalence and the built environment across all four regions, despite varying city and neighborhood values. These high associations between features of the built environment and obesity were achieved without labeling the satellite images or fine-tuning the CNN model to differentiate between images from neighborhoods with high versus low obesity prevalence.

We also studied whether a model fitted solely to these features of the built environment could predict obesity prevalence. We used a random sample representing 60% of the data (n=1,017) in model fitting, and the remaining 40% (n=678) in validation. We found that the features of the built environment were highly predictive of obesity prevalence; correlation between the true and predicted obesity prevalence was 81% (Fig. 2 (b)) with a root mean squared error (hereafter referred to as, error) of 4.35. Similarly, correlations between the actual and model predicted values were 75%, 75%, 78% and 85% for Seattle, Los Angeles, San Antonio, and Memphis, individually (Fig. 5). The error rates across the four regions ranged from 3.22 to 4.54, consistent with that observed for the overall dataset.

In some instances, our models tended to underestimate obesity prevalence especially for Seattle, which could potentially be explained by the presence of green spaces in and the low variation of obesity prevalence across most of the region. Overestimation of obesity prevalence could be potentially due to multiple factors. For example, a high-income region might have features that do not encourage physical activity, however, individuals can afford gym memberships and other recreational facilities. These observations therefore suggest that the features of the built environment can be used in combination with other data sources for monitoring obesity prevalence, and these data could be useful for regions with delayed updates on obesity estimates and for programs focused on reducing obesity.

Although, our models are highly predictive of obesity, we wondered if the density of places of interest (such as, restaurants, parks, recreation centers), which have been extensively studied in association with obesity and physical activity, might be a better predictor. Note that places of interests such as, natural features, are also captured in the features extracted from satellite images. To assess whether the density of places of interest were predictive of obesity, we obtained an extensive list of places of interest (n=96) from Google Nearby Search API and used the data as predictors in regression models for obesity prevalence.

The places of interest data explained 46% of the variation in obesity prevalence across all four regions and achieved a 65% correlation and 5.53 error in out-of-sample predictions (Fig. 2 (c) and (d)). The variation explained at the regional level was approximately 30%, 41%, 58% and 73% for Seattle, Los Angeles, San Antonio and Memphis, respectively. Similarly, correlations between the out-of-sample predictions and actual obesity prevalence were 37%, 66%, 54%, and 66% for Seattle, Memphis, Los Angeles and San Antonio, separately. The out-of-sample prediction errors were higher than the values observed when using data representing features of the built environment to predict obesity prevalence; 4.47, 6.70, 5.39 and 4.09 for Seattle, Memphis, Los Angeles and San Antonio, respectively.

The places of interest data were more predictive for regions with higher obesity prevalence (i.e., San Antonio and Memphis) and less so in regions with lower obesity prevalence. Furthermore, some of the most significant predictors could be directly associated with health, diet and exercise (e.g., gyms, spas, restaurants, bakeries, supermarkets, bowling alleys), while others might be linked to other neighborhood characteristics (e.g., natural features, pet stores, RV parks). Also, strictly restricting the data to places of interest associated with health and exercise resulted in poorer results. We note that while the places of interest data were associated with obesity prevalence, the association was less when compared to the correlation observed with the data extracted from satellite images.

One possible explanation of the significant associations between the data on the built environment features and obesity prevalence is that the data might be representative of socioeconomic indicators such as, poverty. To investigate this assertion, we developed regression models to quantify the association between the data on built environment features and neighborhood per capita income. The joint model fitted on data from all regions explained 77% of the variation in per capita income, and out-of-sample predicted values had a 78% correlation with actual per capita income. These values are similar to what was observed for the amount of obesity prevalence variation explained by our approach. The coefficient of variation ($r^2$) was consistent across the regions with a range between 75% and 81%. However, the correlation between the predicted and actual per capita income varied across the four regions, with observed values of 48%, 61%, 76% and 78% for Memphis, Seattle, San Antonio and Los Angeles, respectively. Additionally, per capita income was consistently underestimated across all cities, suggesting that our approach might need to be refined to produce better predictions of per capita income. Furthermore, the extracted indicators of the built environment might capture additional information not linked to socioeconomic indicators.

**DISCUSSION**

In this study, we used deep learning to extract data representing features of the built environment from high resolution satellite images to examine the association between the built environment and the prevalence of obesity across six cities. The CNN features capture different aspects of the environment such as, greenery and different housing types, that have been associated with physical activity and obesity. Our results demonstrate a consistent association between the built environment and obesity prevalence across neighborhoods with low and high adult obesity prevalence. Although these findings are likely to be explained at least in part by socioeconomic indicators such as, income, our results suggest that the built environment features more consistently predict obesity than per capita income across all regions. A possible explanation is that the features extracted include both man-made changes to the built environment and natural features (e.g., parks and forests) that might not always be predictive or particularly associated with socioeconomic status.

We also demonstrate that the density of places of interest can also predict obesity prevalence, but to a lesser extent when compared to results obtained using the data extracted from satellite images.

While the density of places of interest such as, restaurants, supermarkets and gyms have been studied in association with obesity prevalence, our findings suggest that other places of interest that are not directly linked to food and physical exercise (such as, places of worship or pet shops) might also be strong predictors of obesity prevalence. Additional research is needed to further investigate how these different types of places of interest may be indirectly associated with physical activity or obesity.

Obese persons are at risk for several diseases including, cardiovascular diseases, type 2 diabetes and certain types of cancers. They also experience an overall reduced quality of life when compared to the normal population (2, 42). Our findings are relevant to researchers seeking to develop low cost and timely methods that allow for direct measurement of the built environment to study its association with obesity and other health outcomes. All the data and computational methods used in this study are openly available allowing for comparisons of study results across regions with varying populations and geographies. Additionally, our results are also relevant to people monitoring obesity prevalence or working to develop public health programs to decrease obesity. We show that models fitted solely to the features of the built environment can provide reasonable estimates of neighborhood obesity prevalence, which are typically delayed from official sources by several years. For this study, we used obesity prevalence estimates from 2014, because more recent values are unavailable. Ideally, methods and programs should focus on combining both individual-level and neighborhood-level data to provide timely estimates of neighborhood obesity prevalence.

This study has some limitations. First, the obesity prevalence estimates are based on self-reported height and weight, which have been shown to be biased, and tend to lead to lower estimates of obesity prevalence (43, 44). In addition, BMI does not allow for the direct measurement of body fat, which can vary across sex, age, race and ethnicity. Additionally, mortality and morbidity risk may vary across different race and ethnicities at the same BMI (45, 46). Another limitation of our work is a failure to link specific features of the built environment to high obesity prevalence. To make our approach useful for public health and community planning efforts, our future efforts will focus on deconstructing the results of the CNN to identify and link specific features to obesity prevalence and assess disparities based on neighborhood racial composition and socioeconomic status. Such deconstruction of CNN features is possible with more fine-grained obesity prevalence estimates, that can be used for fine-tuning the neural network. Moreover, Google Street View images can be integrated into the model for drawing inferences with the availability of such data. Socioeconomic status has been linked to obesity and other health outcomes. For example, Gordon-Larsen et al. (2006) found that lower socioeconomic status blocks were more likely to have fewer physical activity facilities and these disparities in access could be associated with differences in overweight patterns (21). Review studies focused on the African American population and disadvantaged populations highlighted a strong association between safety and physical activity, and obesity especially in urban regions (41, 47). These populations were also less likely to have access to needed resources and environments typically associated with lower obesity prevalence. Furthermore, longitudinal studies using our approach can assess changes in the built environment that may be associated with increases or decreases in neighborhood obesity prevalence.

Results in this study strongly support the association between features of the built environment and obesity prevalence. Neighborhood-level interventions to encourage physical activity and increase access to healthy food outlets, could be combined with individual-level interventions to aid in curbing the obesity epidemic.

**MATERIALS AND METHODS**

*Data on Obesity Prevalence*
We obtained 2014 estimates on annual crude obesity prevalence from the 500 cities project (48). We used crude obesity prevalence estimates from 2014 because adjusted estimates and more recent annual estimates were unavailable. These obesity estimates are derived from data from the United States Centers for Disease Control and Prevention (CDC) Behavioral Risk Factor Surveillance System (BRFSS) (2). BRFSS is an annual cross-sectional telephone survey used to measure behavioral risk factors of U.S. residents. The BRFSS includes all 50 U.S. states, the District of Columbia and U.S. territories (i.e., Puerto Rico, Guam and the Virgin Islands).

An individual is considered obese if their body mass index (BMI) is greater than or equal to 30 kg/m$^2$, where BMI is estimated from self-reported weight and height. Certain exclusions were applied, including removal of pregnant women and individuals reporting certain height and weight measurements (49). The data includes respondents 18 years of age and older.

To illustrate our approach, we focused on estimates of obesity prevalence for census tracts in six cities – Los Angeles, California; Memphis, Tennessee; San Antonio, Texas; and Seattle, Tacoma and Bellevue, Washington. Because Seattle, Tacoma and Bellevue are neighboring cities with few census tracts, we combined their data into a single dataset, referred to as, Seattle. These cities were selected to reflect regions with varying obesity prevalence. Recent rankings of obesity prevalence by states lists Tennessee and Texas as sixth and eighth of fifty most obese states, respectively (5). In contrast, the states of Washington and California have lower obesity rates and are ranked thirty-second and forty-seventh of fifty, respectively.

*Satellite Imagery and Places of Interest Data*

We downloaded recent satellite images for each census tract at the zoom level of 18 from the freely available Google Static Maps API. Historical data matching the time period of the obesity estimates was unavailable. The satellite imagery data consisted of approximately 150,000 images. We also obtained a comprehensive list of places of interest (e.g., parks, restaurants, liquor stores, bus stations, night clubs) by performing a nearby search for each location in a square grid spanning a census tract, using the Google Nearby Search API. We included all categories of places of interest available through the API instead of focusing only on physical activity facilities, food and health locations which have been widely studied, because we reasoned that other categories could influence an individual's health behavior and physical activity frequency. For example, a high density of pet stores could indicate high pet ownership which could influence how often people go to parks and take walks around the neighborhood. Furthermore, the places of interest data varied by city, with some cities, such as Los Angeles, having finer classifications than others. The data was further cleaned to avoid duplicate counts of the same location. A comprehensive list of places of interest categories are included in the SI Appendix.

*Deep Neural Network Model*

Data extracted from satellite images have been used in several health-related applications ranging from infectious disease monitoring to estimating socioeconomic indicators such as, poverty (50–52). To extract information from the 150,000 satellite images, we used a convolutional neural network (CNN), which is the state-of-art method for most computer vision tasks such as object recognition, scene labelling, image segmentation and pose estimation. More recently, CNNs have also been trained in image recognition for skin cancer, diagnosis of plant diseases and classification of urban landscapes (53–55).

To train a CNN from scratch to differentiate between regions with low and high obesity prevalence, we need a large corpus consisting of millions of labelled images. However, such training data is unavailable. Instead, we use a transfer learning approach, which involves applying a previously trained network to our dataset of unlabeled images to make inferences. We used a pre-trained network, VGG-F, which has been trained for object recognition on the ImageNet database and is freely available to the research community (56). The convolutional neural network consists of four convolutional layers and four fully connected layers. Each convolutional layer is composed of several two-dimensional filters which activate the features required for classifying an object correctly. During training, the neural network learns to extract gradients, edges and patterns that aid in accurate object detection. The fully connected layers further process these features and convert them into single dimensional vectors. The output layer (final fully connected layer) is originally designed for classifying between 1000 object categories. Essentially, this neural network transforms a large two-dimensional image into a single vector of fixed dimension, containing only the most important descriptors of the image. This feature vector is extracted by deploying the network and making a forward pass through it for each image. It has been shown that these descriptors can be used with linear classifiers or regression models to perform tasks that are much different from object recognition (57). We employ this technique to transform satellite images of dimension 400x400 into a feature vector of length 4096, taken from the seventh hidden layer or third fully-connected layer of the VGG-F network. Further, the mean feature vector for a census tract is computed by taking the average of the vectors for all satellite images belonging to that particular census tract.

*Regression Modeling*

We conducted three different sets of analyses. Our first aim was to quantify the association between the features of the built environment and obesity prevalence at the census tract level. We assessed how much of the variation in obesity prevalence across all census tracts is explained by features of the built environment extracted from satellite images. Since the data contained a large number of features (n=4,096), we applied Elastic Net – a regularization and variable selection technique. A major benefit of Elastic Net is that it combines the advantages of Ridge regression and Least Absolute Shrinkage and Selection Operator (LASSO); insignificant covariates are eliminated, while correlated variables that are significant are maintained (58). Next, we evaluated how well our model predicts obesity prevalence across all cities by splitting the data into two sets – a random sample representing sixty percent of the data for fitting and the remaining forty percent for model evaluation.

Our second aim was to quantify the association between the neighborhood density of places of interest (such as, fast food outlets and recreational facilities) to obesity prevalence. We used the same process for variable selection and regression modeling as previously described. We then compared the model coefficient of determination ($r^2$), root mean squared error (RMSE) and Pearson correlation between the actual and predicted observations to our previous results on predicting obesity using data on the built environment extracted from satellite images.

Lastly, we fit separate models to quantify how well the features of the built environment extracted from satellite images predict socioeconomic variables, such as per capita income. This analysis was undertaken because obesity prevalence tends to correlate with socioeconomic status and we wanted to investigate if this could partially explain the performance of our models in predicting obesity prevalence.

We used the R statistical software for all the regression modeling. Five-fold cross validation was applied to models for all regions except Memphis, for which we used a three-fold cross validation because the sample size was less than 200, which limits the number of data points used in each fold. To prevent over-fitting, we also limited the number of selected features to be less than or equal to the number of census of tracts.

**FIGURES**

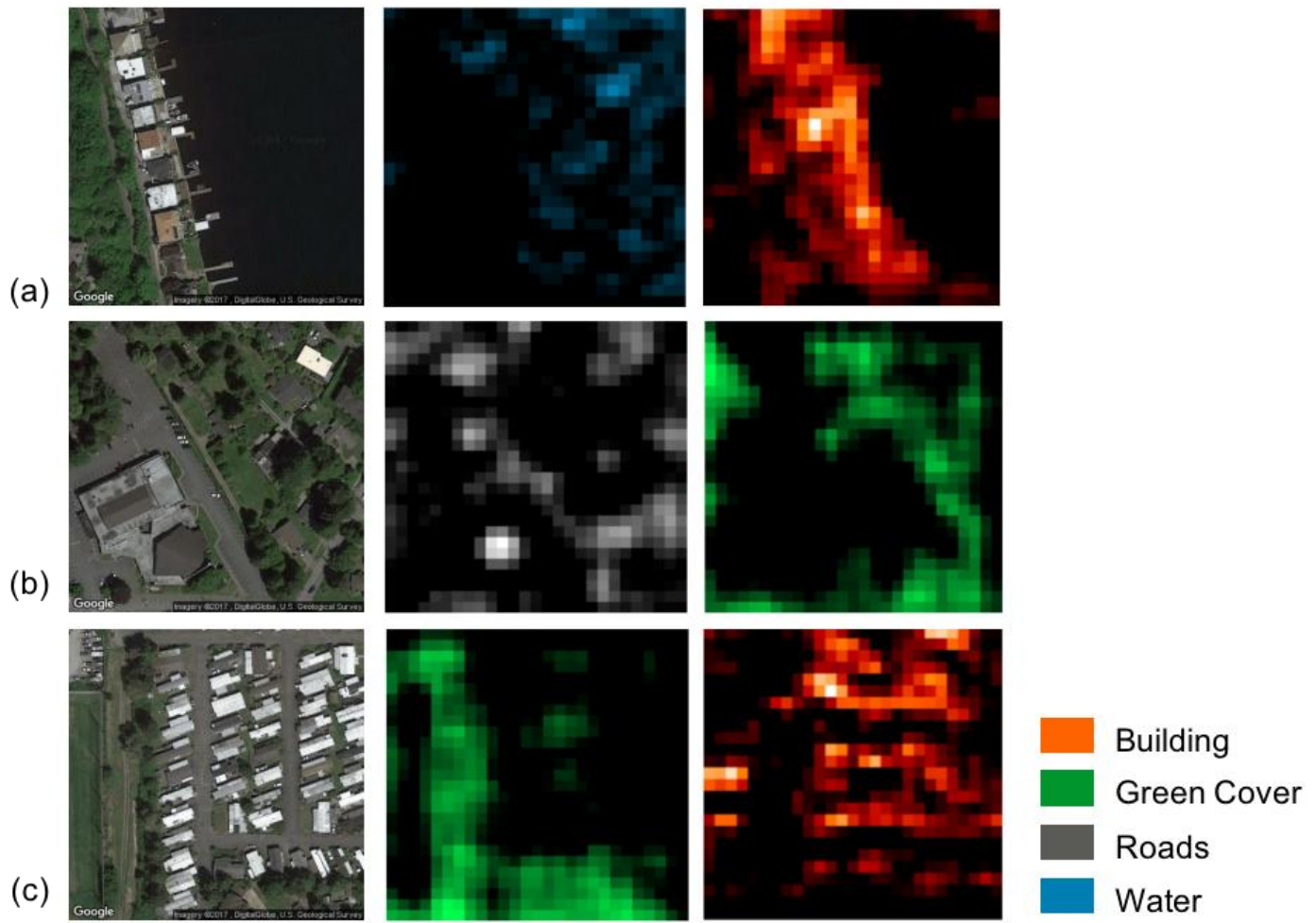

**Fig. 1. Illustration of features identified by neural network model.** The images on the left column are satellite images taken from Google Static Maps API. Images in other columns are activation maps taken from the second convolutional layer of VGG-F network after forward pass of the respective satellite images through the network. We can see that the neural network

understands image by interpreting the output from filters learned during the training phase. The filter outputs correspond to water and buildings in row (a), roads and green cover in row (b), green cover and buildings in row (c). The activation maps may not always align exactly with the original image due to padding of output within the network.

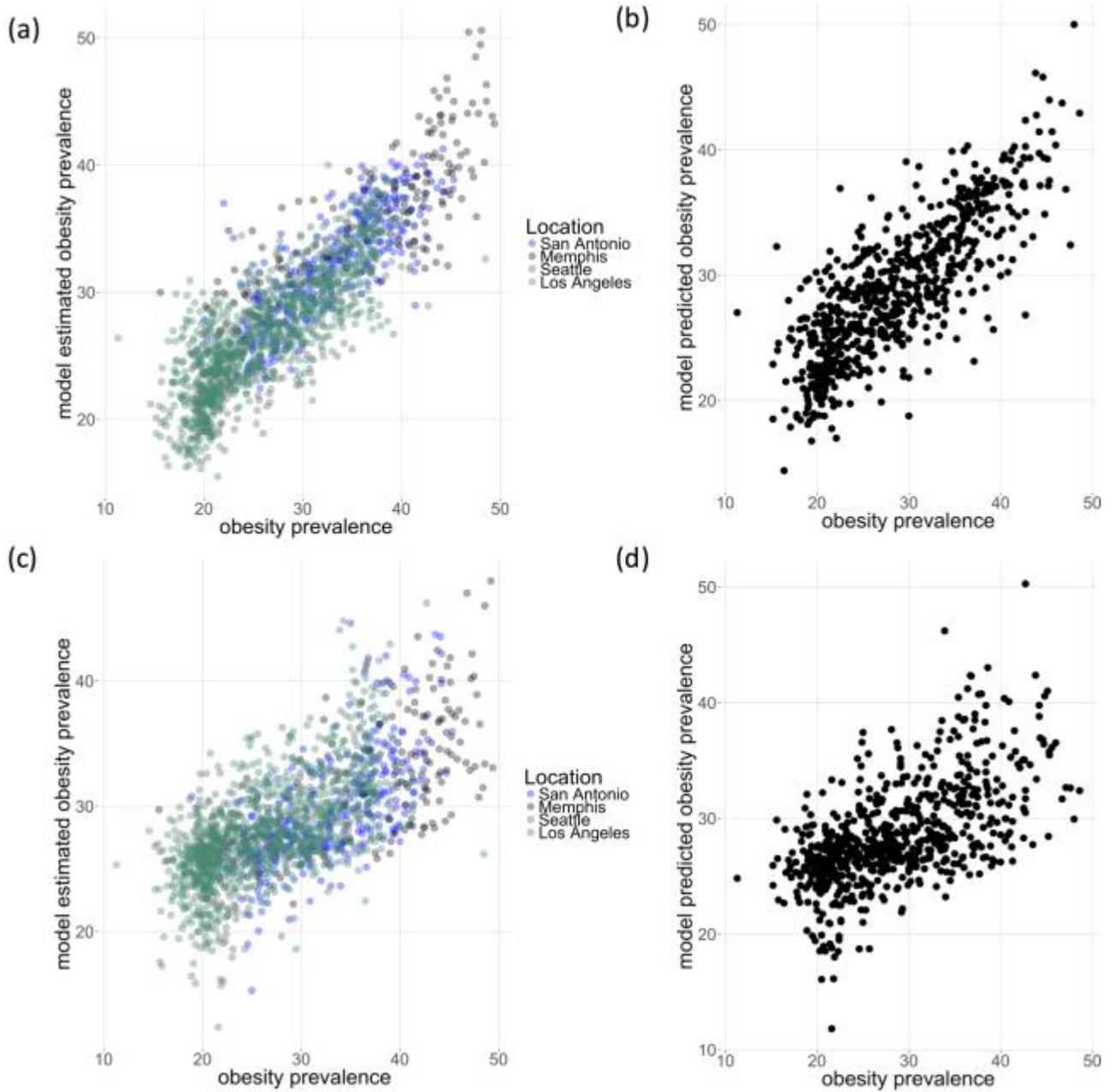

**Fig. 2. Scatterplots of model estimated and predicted obesity prevalence plotted against actual obesity prevalence.** (a) cross-validated model estimates and (b) out-of-sample predictions of obesity prevalence based on features of the built environment extracted from satellite images. (c) cross-validated model estimates and (d) out-of-sample predictions of obesity prevalence based on the density of places of interest data.

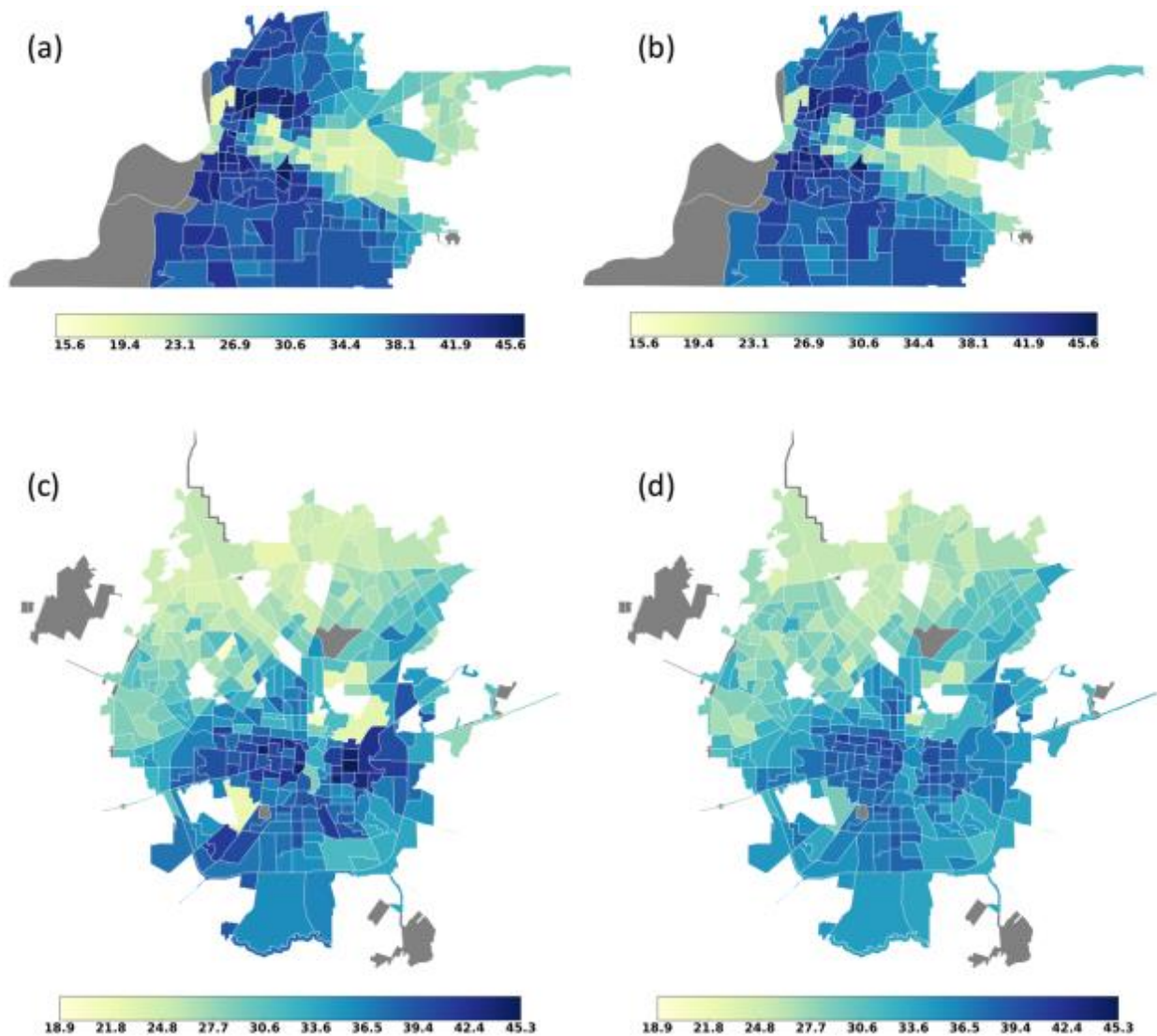

**Fig. 3. Actual obesity prevalence and cross-validated model estimates of obesity prevalence.** (a) actual and (b) cross-validated estimates of obesity prevalence for Memphis, Tennessee based on features of the built environment extracted from satellite images. (c) actual and (d) cross-

validated estimates of obesity prevalence for San Antonio, Texas based on features of the built environment extracted from satellite images. The gray shaded regions do not have data.

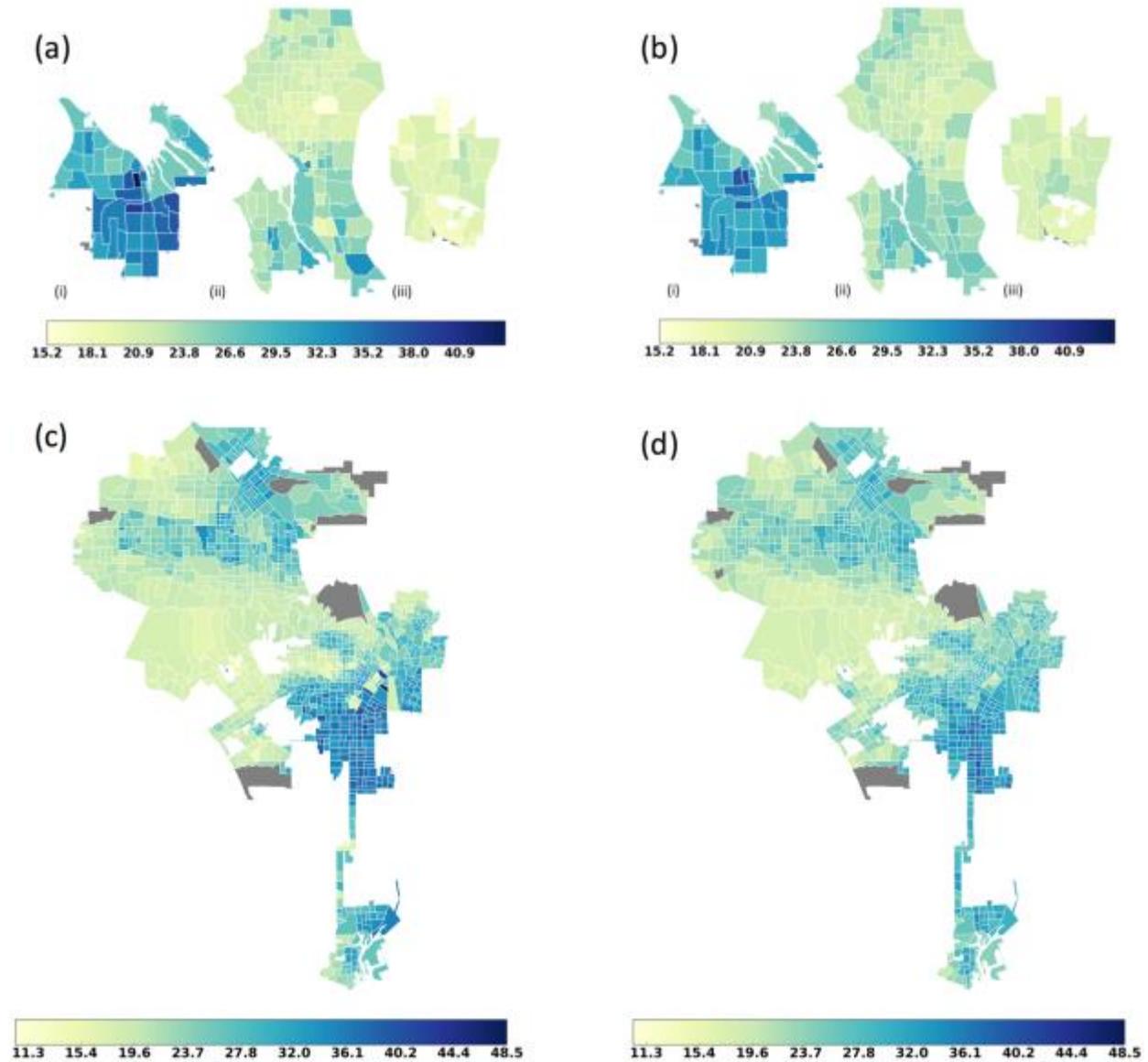

**Fig. 4. Actual obesity prevalence and cross-validated model estimates of obesity prevalence.** (a) actual and (b) cross-validated estimates of obesity prevalence for Bellevue (i), Seattle (ii) and Tacoma (iii), Washington based on features of the built environment extracted from satellite

images. (c) actual and (d) cross-validated estimates of obesity prevalence for Los Angeles, California based on features of the built environment extracted from satellite images. The gray shaded regions do not have data.

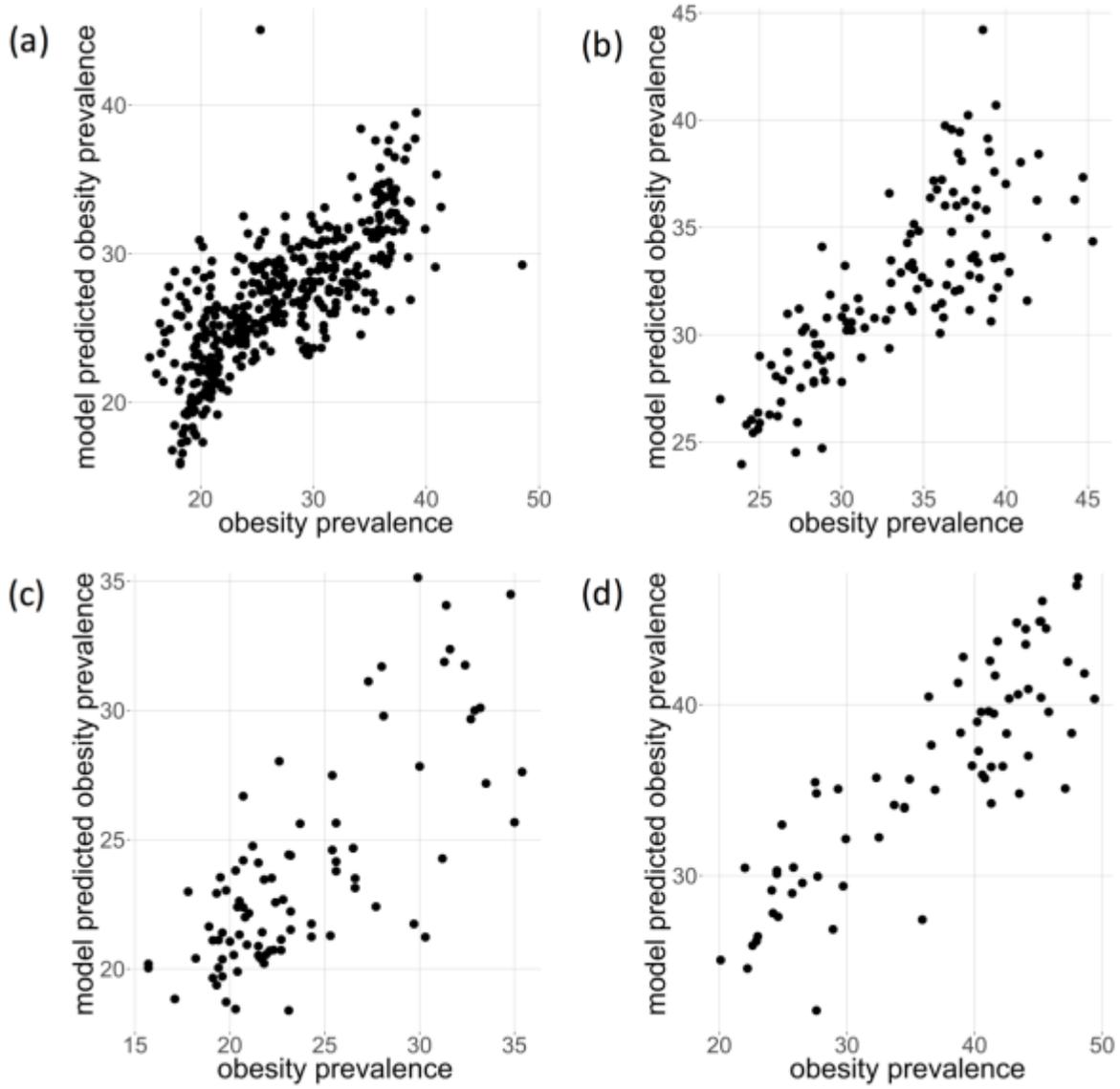

**Fig. 5. Actual obesity prevalence plotted against model predictions.** (a) Los Angeles (b) San Antonio, (c) Seattle and (d) Memphis.